\documentclass[amsmath,trackchanges]{aastex702}
\usepackage[scheme=plain, UTF8]{ctex}
\begin{document}

\title{Collisional excitation of SiC$_6$ by  helium: cross sections, rate coefficients, and astrophysical implications}

\author[orcid=0009-0009-3771-8263, gname=Liman, sname=Tian]{Liman Tian(田丽嫚)}
\affiliation{College of Optical, Mechanical and Electrical Engineering (College of Intelligent Equipment Engineering), Zhejiang A \& F University, Hangzhou 311300, People's Republic of China}
\email{tianlm2258@163.com}

\author[orcid=0009-0003-5090-9618, gname=Ziyu, sname=Xiang]{Ziyu Xiang(项梓瑜)}
\affiliation{College of Optical, Mechanical and Electrical Engineering (College of Intelligent Equipment Engineering), Zhejiang A \& F University, Hangzhou 311300, People's Republic of China}
\email{xiangzy2609@163.com}

\author[orcid=0000-0002-3158-1253, gname=Chuangang, sname=Ning]{Chuangang Ning(宁传刚)}
\affiliation{Department of Physics, Tsinghua University, Beijing 100084, People's Republic of China}
\email{ningcg@tsinghua.edu.cn}

\author[orcid=0009-0001-2007-2007, gname=Xiaolin, sname=Chen]{Xiaolin Chen(陈孝林)}
\affiliation{College of Optical, Mechanical and Electrical Engineering (College of Intelligent Equipment Engineering), Zhejiang A \& F University, Hangzhou 311300, People's Republic of China}
\email[show]{chenxl13@zafu.edu.cn}
\correspondingauthor{Xiaolin Chen}


\begin{abstract}
{Rotational transitions of SiC$_6$ with $j=26$--$40$ have been detected in IRC+10216, but collisional rate coefficients for SiC$_6$ with He or H$_2$ were previously unavailable. We present rate coefficients for SiC$_6$ rotational excitation by He, calculated using a two-dimensional potential energy surface (PES) at the RCCSD(T)-F12 level under the rigid-rotor approximation. Close-coupling (CC) calculations were performed for low rotational levels, while the coupled-states (CS) method was used for higher levels. Convergence tests were conducted, and CC and CS scattering cross sections were compared. The CS rate coefficients cover rotational levels up to $j=40$ and temperatures of 3--60\,K. Although CC and CS rate coefficients differ by up to a factor of $\sim 2$ for some transitions, particularly those affected by weak coupling and resonances, they yield nearly identical excitation and brightness temperatures for the high-$j$ transitions examined, including $19\rightarrow18$. These results support the reliability of the CS method for non-LTE modelling of the high-$j$ transitions considered, for which  CC calculations are computationally impractical. We use reduced-mass scaling of the SiC$_6$--He rate coefficients to estimate those for SiC$_6$--H$_2$ and model the observed transitions. A $\chi^2$ analysis of the observed line fluxes, excluding heavily blended features, yields a best-fitting of $n(\mathrm{H_2})=4.33\times10^5$\,cm$^{-3}$ and  $N(\mathrm{SiC_6})=8.11\times10^{11}$\,cm$^{-2}$. The latter agrees excellently with the previous observational estimate of $(7.9\pm1.5)\times10^{11}$\,cm$^{-2}$, supporting the consistency of our non-LTE modelling with observational constraints.}

\end{abstract}



\section{Introduction}

A wide variety of molecules exist in the interstellar medium (ISM) and the circumstellar envelopes of evolved stars \citep{Endres2016}; among them, numerous silicon-bearing species have been detected, including simple molecules such as SiH$_4$, SiN, SiS and SiP \citep{Morris1975,Goldhaber1984,Turner1992,Koelemay2022}. The chemical behaviour of silicon within the ISM is a crucial aspect of understanding dust formation and evolution. Although the abundance of gas-phase silicon is significantly depleted due to its incorporation into silicate dust, silicon-bearing molecules still exhibit complex chemical networks in specific environments, such as carbon-rich circumstellar envelopes \citep{MacKay1999}.  Linear silicon-carbon chains, SiC$_n$, are of particular interest because their rotational spectra provide observable tracers of silicon-bearing carbon-chain chemistry. Interpreting these spectra quantitatively, however, requires reliable molecular spectroscopic data and collisional excitation rates.

The history of laboratory and astronomical detection of silicon-carbon chain molecules spans more than four decades. In 1984, Thaddeus \textit{et al}. first identified cyclic SiC$_2$ in IRC\,+10216, marking the first detection of a cyclic molecule in interstellar space \citep{Thaddeus1984}. Subsequently, linear SiC and SiC$_2$ were characterized in the laboratory \citep{Verma1974,Cernicharo1989}. However, the detection of longer silicon-carbon species was long hindered by a lack of laboratory spectroscopic data. A breakthrough occurred in 2000 when McCarthy \textit{et al}., using Fourier-transform microwave spectroscopy on supersonic molecular beams, detected five new linear silicon-carbon chain molecules---SiC$_3$, SiC$_5$, SiC$_6$, SiC$_7$, and SiC$_8$---for the first time; they confirmed that even-numbered chain members possess an $\mathrm{X}^{1}\Sigma^{+}$ electronic ground state, while odd-numbered members possess an $\mathrm{X}^{3}\Sigma^{-}$ state \citep{McCarthy2000}. This systematic laboratory work laid the spectroscopic foundation for subsequent radio astronomical searches.

In terms of astronomical observations, the silicon-carbon chemical network of IRC\,+10216 has been progressively revealed. SiC$_4$ was first detected in this source by \citet{Ohishi1989}, with a column density approximately one-tenth that of SiC and one-thirtieth that of c-SiC$_2$ \citep{Thaddeus1984,Cernicharo1989}. Subsequently, longer-chain silicon-carbon species were successively identified. Recent sensitive Q-band observations with the Yebes 40-m telescope have further expanded the known SiC$_n$ inventory in IRC+10216.  \citet{Cernicharo2025} reported the detection of linear SiC$_3$ and SiC$_5$, while  \citet{Pardo2025}  reported the detection of SiC$_6$ through a series of rotational transitions. They detected 14 SiC$_6$ transitions with $J_{\rm up}=26-40$, including nine isolated or only weakly blended features, and derived a column density of approximately $8\times10^{11}\,\mathrm{cm^{-2}}$ and a rotational temperature of about 19~K. This detection provides a direct observational motivation for determining the collisional excitation of SiC$_6$. The rotational temperature derived from these observations does not, on its own, establish whether the emission arises under local thermodynamic equilibrium (LTE) or non-LTE conditions. Instead, it provides an observational constraint against which multi-level radiative-transfer models can be tested. In particular, the high-$j$ transitions detected by \citet{Pardo2025} probe rotational levels for which reliable collisional data are essential for assessing the excitation conditions.

Research on the collisional excitation of silicon-carbon chain molecules is currently in its infancy. Collision rate coefficients for c-SiC$_3$--He were reported for the first time only recently \citep{Mehnen2026}, and data on the rotational excitation of SiC$_4$--He have also just become available \citep{Chefai2024}. These recent studies demonstrate the growing importance of accurate molecular collision data for interpreting astronomical observations of silicon-carbon species. In contrast, to the best of our knowledge, there is no previously published state-to-state potential energy surface (PES) or rotationally inelastic rate-coefficient dataset for collisions of SiC$_6$ with He or H$_2$. This lack of data limits the ability to perform quantitative non-LTE modelling of the recently detected SiC$_6$ emission.

This study aims to construct a PES using a high-level correlated \textit{ab initio} method and to provide rotationally inelastic rate coefficients for the SiC$_6$--He collisional system by combining close-coupling (CC) and coupled-states (CS) quantum scattering calculations. A particular emphasis is placed on assessing the differences between CC and CS results and evaluating the applicability of the CS approximation to the rotational excitation of SiC$_6$. Subsequently, radiative-transfer calculations are employed to investigate the excitation of SiC$_6$ under representative low-density conditions and to assess the consequences of using CS rate coefficients for the high-$j$ transitions relevant to the recent detection. In this study, we establish the first state-to-state rotationally inelastic collisional dataset for SiC$_6$--He. The paper is organized as follows. Section 2 describes the quantum scattering calculations, including the construction and analytical representation of the SiC$_6$--He PES. Section 3 presents the convergence tests and the comparison between CC and CS scattering calculations, followed by the resulting rate coefficients for rotational levels up to $j=40$. Section 4 discusses the astrophysical applications to SiC$_6$ excitation. Section 5 summarizes the main conclusions.

\section{The \textit{Ab Initio} Computed Interaction Potential}
\label{sec:scattering}
\subsection{Potential energy surface}

A two-dimensional interaction potential energy surface (PES) for the SiC$_6$--He system was constructed using the rigid-rotor approximation, in which bond lengths are fixed and the intermolecular geometry is described by the separation and orientation of He relative to SiC$_6$. For SiC$_6$ ($\omega_e = 56$\,cm$^{-1}$; \citep{Gordon2000}), the thermal population of vibrationally excited states is expected to be small at the low interstellar temperatures relevant to this study; for example, the Boltzmann factor for the first vibrationally excited state at 20\,K is $\exp[-56/(k_{\mathrm B}T)]\approx 1.8\%$.
This approximation has also been widely applied to related systems, such as SiC$_4$--He \citep{Chefai2024}, HNSi/HSiN--He \citep{Tian2026}, and CaO/CaS--He \citep{Tian20261}.


The interaction PES is described using Jacobi coordinates $(R,\theta)$, where $R$ is the distance from He to the centre of mass of SiC$_6$, and $\theta=0^{\circ}$ corresponds to the He--C$_6$Si arrangement. The bond lengths are $r_{\mathrm{C}_1\mathrm{C}_2}=1.277\,\text{\AA}$, $r_{\mathrm{C}_2\mathrm{C}_3}=1.300\,\text{\AA}$,
$r_{\mathrm{C}_3\mathrm{C}_4}=1.256\,\text{\AA}$,
$r_{\mathrm{C}_4\mathrm{C}_5}=1.290\,\text{\AA}$,
$r_{\mathrm{C}_5\mathrm{C}_6}=1.282\,\text{\AA}$,
$r_{\mathrm{C}_6\mathrm{Si}}=1.689\,\text{\AA}$.
All the SiC$_6$ bond distances are were fixed at the equilibrium values for the ground electronic state (X$^{1}\Sigma^{+}$) reported by \citet{Gordon2000}.
In this study, a two-dimensional interaction PES  for the SiC$_6$(X$^{1}\Sigma^{+}$)--He($^{1}$S) complex was constructed using the RCCSD(T)-F12A/aug-cc-pVTZ method \citep{Knizia2009}. The radial coordinate $R$ ranged from 2.0 to 20.0\,Å with a variable step size (as fine as 0.05\,Å), while the angle $\theta$ was sampled from $0^{\circ}$ to $180^{\circ}$ at $2.5^{\circ}$ intervals; a total of 7279 geometric configurations were calculated across the sampled intermolecular-distance and angular ranges. The interaction energy was calculated using the Boys--Bernardi counterpoise correction procedure to account for basis-set superposition error (BSSE) \citep{Boys1970}:
\begin{equation}
	\label{eq.1}
	\ V(R,\theta )={{E}_{\text{Si}{{\text{C}}_{\text{6}}}\text{--He}}}(R,\theta )-{{E}_{\text{Si}{{\text{C}}_{\text{6}}}}}(R,\theta )-{{E}_{\text{He}}}(R,\theta )\
\end{equation}
where $E_{\mathrm{SiC_6-He}}$, $E_{\mathrm{SiC_6}}$, and $E_{\mathrm{He}}$ are the energies of the SiC$_6$--He system, the SiC$_6$ molecule, and the He atom, respectively.

\begin{figure}[ht!]
	\centering
	\includegraphics[width=\hsize]{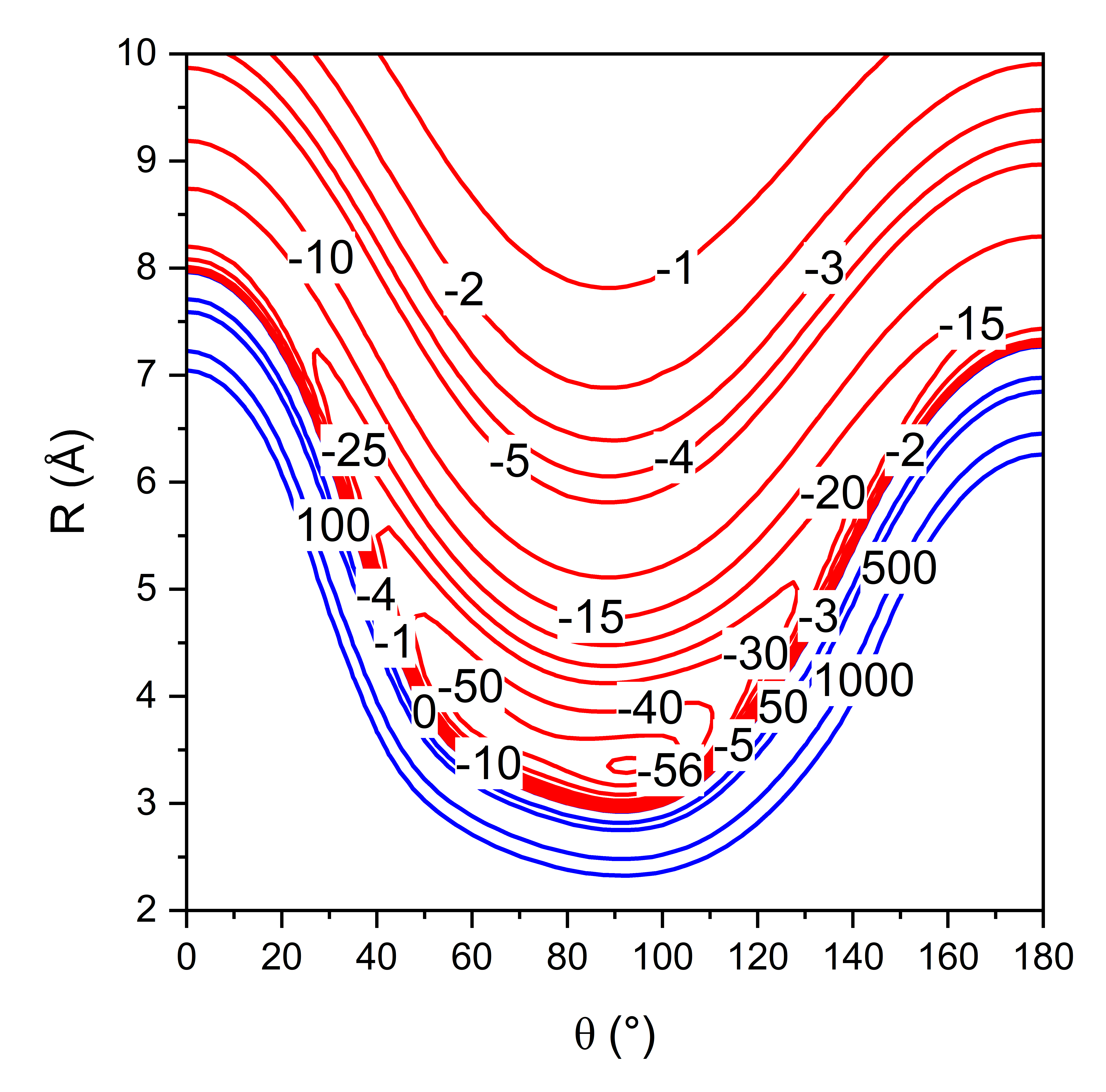}
	\caption{Contour plot (in cm$^{-1}$) of the SiC$_6$--He PES as a function of Jacobi coordinates $R$ and $\theta$. Negative values are shown in red and positive values in blue.}
	\label{fig.1}
\end{figure}

The resulting PES contour plot is shown in Fig.~\ref{fig.1}; the calculated PES has  a global minimum at $R = 3.35$\,Å and $\theta = 92.5^{\circ}$, with a potential well depth of 56.75\,cm$^{-1}$. This well depth is similar to that of its homologue, SiC$_4$ (54.21\,cm$^{-1}$). The Mulliken charge analysis shows that C$_4$ carries the largest positive charge (+1.32465) among all atoms, and the global minimum of the PES is located at $\theta = 92.5^\circ$, corresponding to a perpendicular approach of He toward the molecular axis near C$_4$; this correlation between the charge distribution and the minimum-energy geometry supports the reliability of the computed PES. All \textit{ab initio} calculations in this study were performed using the Molpro software \citep{Werner2011}.

\subsection{Analytical Representation}
It should be noted that for long, linear structures like the SiC$_6$ molecule, the PES exhibits numerous ``ghost points`` with extremely high energies at short ranges. If these points are used directly for Legendre polynomial fitting, the large values in the high-energy region dominate the fit, leading to reduced accuracy in the low-energy region that is most relevant to scattering calculations or, in severe cases, numerical instability.
To address this issue, we employ a simple scaling method proposed by \citet{Gonzalez2024}---an approach previously applied to similar systems \citep{Sahnoun2018,Khadri2020}. By setting an appropriate potential energy threshold to ensure a smooth transition between low- and high-energy regimes, this method mitigates the influence of extremely repulsive points on the fitting procedure while preserving the original potential values below the lower threshold. The computational workflow is outlined below:	
\begin{equation}
	\label{eq.2}
	V_s =
	\begin{cases}
		V_a, & V_a \le V_{T1}, \\[4pt]
		V_{T1} + (V_{T2}-V_{T1})\,S\!\left(\dfrac{V_a-V_{T1}}{V_{T2}-V_{T1}}\right),
		& V_{T1} < V_a \le V_{T2}, \\[8pt]
		V_{T1} + \dfrac{2}{\pi}(V_{T2}-V_{T1}), & V_a > V_{T2}.
	\end{cases}
\end{equation}

In the above expression, $V_s$ represents the scaled potential energy, and $V_a$ is the \textit{ab initio} energy. $V_{T1}$ and $V_{T2}$ are taken as 1000 and 10000\,cm$^{-1}$, respectively, for the present study. The transition among the three scaling windows is controlled by a function $S(x)=\frac{2}{\pi}\sin\!\left(\frac{\pi}{2}x\right)$, which ensures continuity of the scaled potential and its first derivative at both thresholds.

To characterize the SiC$_6$--He interaction potential, we adopted the standard Legendre expansion reported by \citet{Werner1988} and fitted the original two-dimensional mesh data obtained from \textit{ab initio} calculations:
\begin{equation}
	\label{eq.3}
	\ V(R,\theta )=\sum\limits_{\lambda =0}^{{{\lambda }_{\max }}}{{{V}_{\lambda }}}(R){{P}_{\lambda }}(\cos \theta ) \
\end{equation}
where $P_\lambda(\cos\theta)$ is the Legendre polynomial of order $\lambda$. We expanded the order up to $\lambda_{\max} = 72$, obtaining a fitting error of 0.5\% according to the error metric defined for the PES fit. The results indicate that among all the anisotropic expansion terms ($\lambda>0$), the $V_2$ term (corresponding to $\lambda=2$) exhibits the largest amplitude. This term contributes directly to coupling between rotational channels with $\Delta j=0, \pm 2$, subject to the angular-momentum coupling rules \citep{McCurdy1977}; consequently, its relatively large amplitude can make it an important contributor to the subsequent scattering dynamics.


\section{Scattering calculations}
After the PES fitting was completed, we calculated the collision cross sections $\sigma_{j\rightarrow j'}(E_k)$ using the CC and CS methods, respectively; (the CC method covered total energies up to 200\,cm$^{-1}$, whereas the CS method extended the range to 500\,cm$^{-1}$). The rotational energy levels of SiC$_6$ are given by the following expression:
\begin{equation}
	\label{eq.4}
	\ E_j = B_0\,j(j+1) - D_0\,[j(j+1)]^2 \,.
\end{equation}

For the SiC$_6$ collision system, the rotational and centrifugal distortion constants reported by \citet{Gordon2000} were used: $B_0 = 0.020389$\,cm$^{-1}$ and $D_0 = 1.83\times10^{-10}$\,cm$^{-1}$; the state-to-state integral cross-sections for transitions from initial rotational energy level $j$ to final level $j'$ were calculated using the standard expression given by \citet{Green1975}.

\begin{equation}
	\label{eq.5}
	\begin{split}
		\sigma_{j \to j'}(E_k) ={}& \frac{\pi}{k_j^2 (2j+1)} \sum_{J=0} \sum_{l=|J-j|}^{J+j} \sum_{l'=|J-j'|}^{J+j'} (2J+1) \\
		&\times \left| \delta_{jj'} \delta_{ll'} - S_{jj'll'}^{J}(E) \right|^2 ,
	\end{split}
\end{equation}

In the equation above, $k_j = \sqrt{2\mu(E-E_j)}/\hbar$ stands for the wave number in the incident channel, $E$ is the total energy of the system, and $E_j$ corresponds to the rotational energy of the initial state. For the SiC$_6$--He collision system, the reduced mass is $\mu = 3.848523$\,amu. The orbital angular-momentum quantum numbers of the incoming and outgoing channels are denoted by $l$ and $l'$, respectively, while $J$ denotes the total angular-momentum quantum number.

In the calculation of scattering cross-sections, we employed both the CC and CS methods. A convergence test of the radial coefficients $V_\lambda(R)$ showed that terms with $\lambda>45$ contribute negligibly to the expansion; therefore, $\lambda_{\max}=45$ was adopted for the scattering calculations. Independently, the rotational basis set truncation parameter $j_{\max}$ was optimized for each method and energy range through separate convergence tests, as described below.
To assess numerical convergence, we tested parameters including the propagation boundaries and step sizes, aiming for changes in the calculated cross sections of less than  5\%. We optimized the rotational basis set truncation parameter $j_{max}$ for different methods and energy ranges; for the CC method, due to computational cost constraints, calculations were converged up to $j \le 19$ with a total energy range extending to 100 cm$^{-1}$, utilizing fine step sizes of 0.05, 0.1, and 1 cm$^{-1}$ to accurately capture resonance features. In the 100--200 cm$^{-1}$ range, total energy points were spaced at 5 cm$^{-1}$ intervals, and $j_{max}$ increased gradually from 32 to 45. For the CS method, to calculate transitions involving rotational levels up to $j=40$, higher $j_{max}$ values were required, increasing from 50 to 80 across the energy range considered.
The radial propagation range was set from $R_{\mathrm{MIN}}=2.0$\,Å to $R_{\mathrm{MAX}}=20.0$\,Å. For energies $E<50$\,cm$^{-1}$, the integration step size was set to 30 steps per de Broglie wavelength. At each collision energy, $J_{\mathrm{tot}}$ was increased until the cross sections converged; JTOTU= 99999 set the upper limit of the built-in JTOT loop. The convergence thresholds for diagonal and off-diagonal elements, DTOL and OTOL, were set to $0.01$\ and $0.001$\, respectively. The coupled equations are solved using the improved log-derivative method proposed by \citet{Manolopoulos1986} in the short-range region and the Airy propagator \citep{Alexander1987} in the long-range region. All scattering calculations were performed using the MOLSCAT software package \citep{Hutson2019}.

\begin{figure}
	\centering
	\includegraphics[width=\hsize]{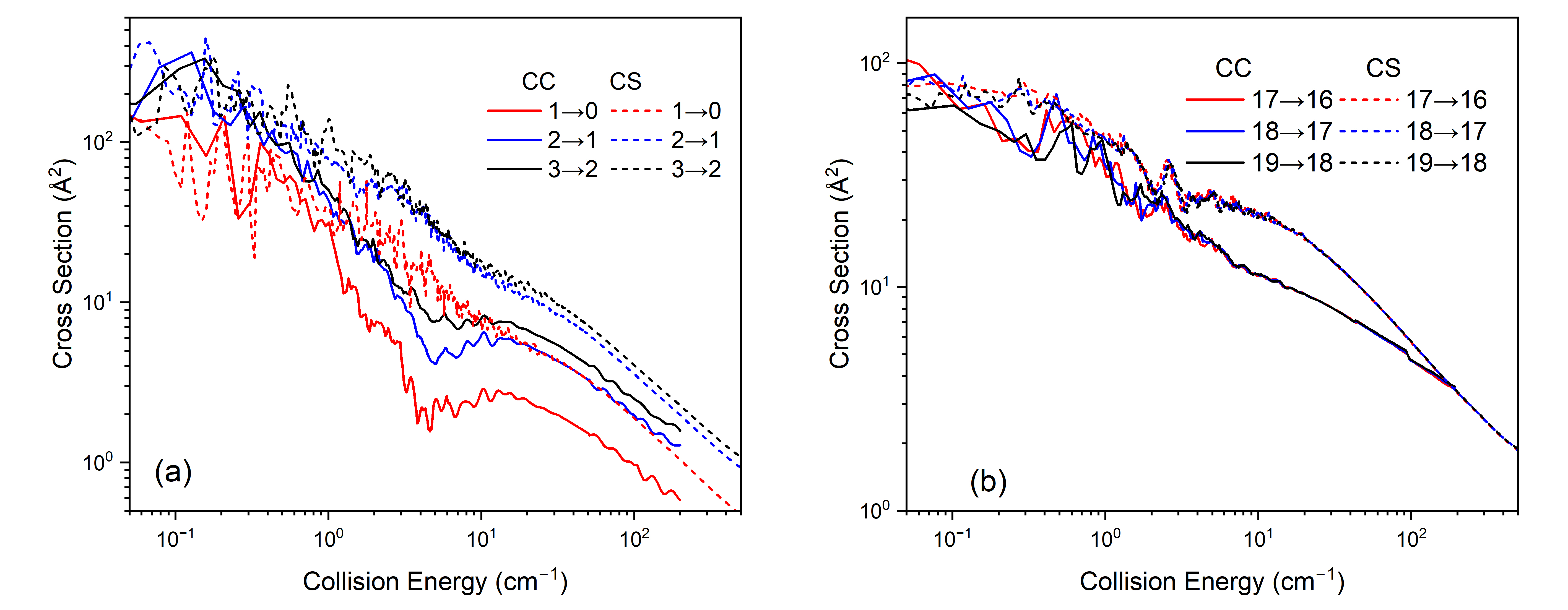}
	\caption{Cross sections calculated using the CC and CS methods as a function of collision
		energy for representative low and high$-j$ transitions.}
	\label{fig.2}
\end{figure}

Because converged CC cross-section calculations involving highly excited rotational levels (up to $j=40$) are computationally demanding in terms of CPU time and memory, we evaluated the feasibility of using the CS approximation for higher rotational levels. We first examined the general trends of cross-sections versus collision energy for both methods. Fig.~\ref{fig.2} illustrates the energy dependence of cross-sections for representative rotational transitions, with low-lying transitions ($1\to0$, $2\to1$, $3\to2$) in panel (a) and higher transitions ($17\to16$, $18\to17$, $19\to18$) in panel (b); solid and dashed lines represent CC and CS results, respectively. Both methods exhibit similar energy-dependent behaviour: dense resonance structures appear at low energies ($E < 10\ \mathrm{cm}^{-1}$), while the cross sections generally decrease with increasing collision energy at higher energies ($E > 10\ \mathrm{cm}^{-1}$). The low-amplitude fluctuations in the CC cross sections at higher energies arise because these calculations are extremely time-consuming; based on the relative convergence of the scattering cross sections, we had to adopt $\mathrm{JSTEP}=2$ or $3$ to accelerate the calculations. A closer comparison reveals that the discrepancy between the two methods depends significantly on the initial rotational level. For low rotational levels (panel a), the CS results are systematically higher than the CC results in the high-energy region, whereas for higher rotational levels (panel b), the two methods agree more closely, with the largest discrepancies occurring in the low-energy resonance region and generally decreasing as the energy increases. This behaviour suggests that the CS approximation can provide a useful description of the selected high-$j$ transitions, particularly at higher collision energies, supporting its use for the high-$j$ calculations for which fully converged CC results are computationally impractical.

\begin{figure*}
	\centering
	\includegraphics[width=\hsize]{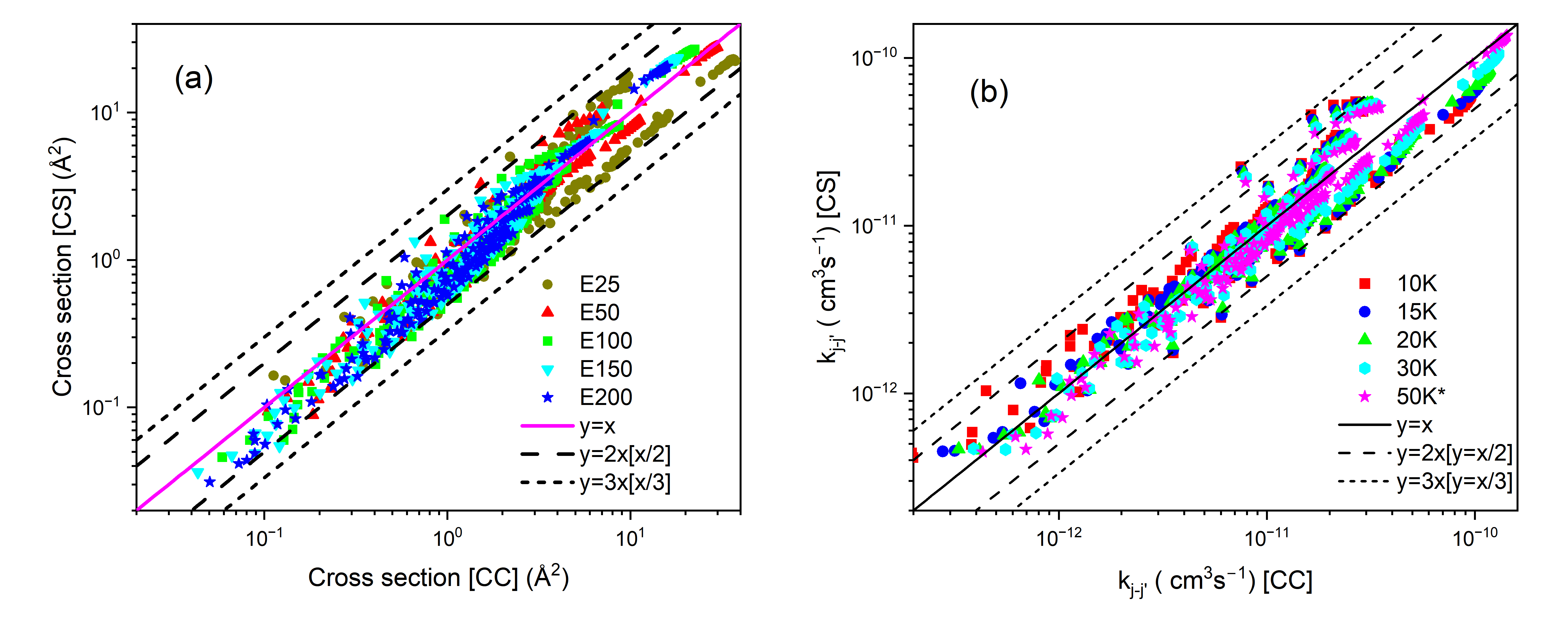}
	\caption{State-to-state rotational de-excitation of SiC$_6$--He calculated using the CC and CS methods. (a) Cross sections compared at total energies of 25, 50, 100, 150 and 200 cm$^{-1}$. (b) Comparison of the thermal rate coefficients.}
	\label{fig.3}
\end{figure*}



To quantify the discrepancy between the CC and CS methods across the available transition channels, we selected five representative total-energy points ($E=25$, 50, 100, 150, and 200\,cm$^{-1}$). As shown in Fig.~\ref{fig.3}(a), we compared the CC and CS rotationally inelastic cross sections for SiC$_6$--He for transitions originating from levels with $j,j'\leq19$. The horizontal axis represents the cross-sections obtained via the CC method, while the vertical axis represents those obtained via the CS method. The cross sections calculated using the CS and CC methods at the five selected energies are distributed generally along the diagonal line $y=x$, indicating overall agreement between the two methods. The cross-section results for the vast majority of transition channels fall within the range of 0.4 to 2.5 times the CC values. As the energy increases from 25 \,cm$^{-1}$ to 200 \,cm$^{-1}$, the agreement between the two sets of calculations generally improves, with most cross sections falling within a factor of two. These results support the use of the CS approximation for the high-energy transitions examined here, where its cross sections show reasonable agreement with the CC results.

The rotational de-excitation rate coefficients were obtained by thermal averaging of the cross sections:
\begin{equation}
	\label{eq.6}
	\ {{k}_{j\to {j}'}}(T)={{\left( \frac{8}{\pi \mu k_{\text{B}}^{3}{{T}^{3}}} \right)}^{1/2}}\int_{0}^{\infty }{{{E}_{k}}}{{\sigma }_{j\to {j}'}}({{E}_{k}}){{e}^{-{{E}_{k}}/{{k}_{\text{B}}}T}}d{{E}_{k}} \
\end{equation}
where $\mu$ is the reduced mass of the SiC$_6$--He collision system, and $k_B$ is the Boltzmann constant.

Building upon the cross-section comparison, we further compared the CC and CS rotational de-excitation rate coefficients for SiC$_6$--He for the transitions included in the CC calculations with $j\leq19$. The comparisons were performed at five temperatures (10, 15, 20, 30, and 50K), and the results are shown in Fig.~\ref{fig.3}(b). The CC rate coefficients at 50 K were obtained by mathematical extrapolation and are marked with an asterisk (*); they are included only for qualitative comparison with the CS results. 
The results show close correspondence between the rate coefficients obtained by the two methods across all five temperatures. Most CS/CC ratios remain within a factor of two, with only a few points approaching or slightly exceeding a factor of three. As the temperature rises from 10\,K to 50 \,K, the dispersion of data points decreases slightly, with the great majority clustering tightly around the 1:1 reference line .These results suggest that the CS approximation provides a reasonable description of the dominant rate coefficients. The larger relative deviations found for some smaller rate coefficients may have a limited effect on the modeled excitation, as assessed in the radiative transfer calculations below. Consequently, the radiative transfer simulations of higher-$j$ transitions in this study utilize rate coefficients derived from the CS method. This approach aligns with the conclusions drawn from CC/CS comparisons for systems such as HC$_5$N, HC$_7$N \citep{Bop2025}, and C$_5$H$_6$ \citep{Demes2026}.

\begin{figure}
	\centering
	\includegraphics[width=\hsize]{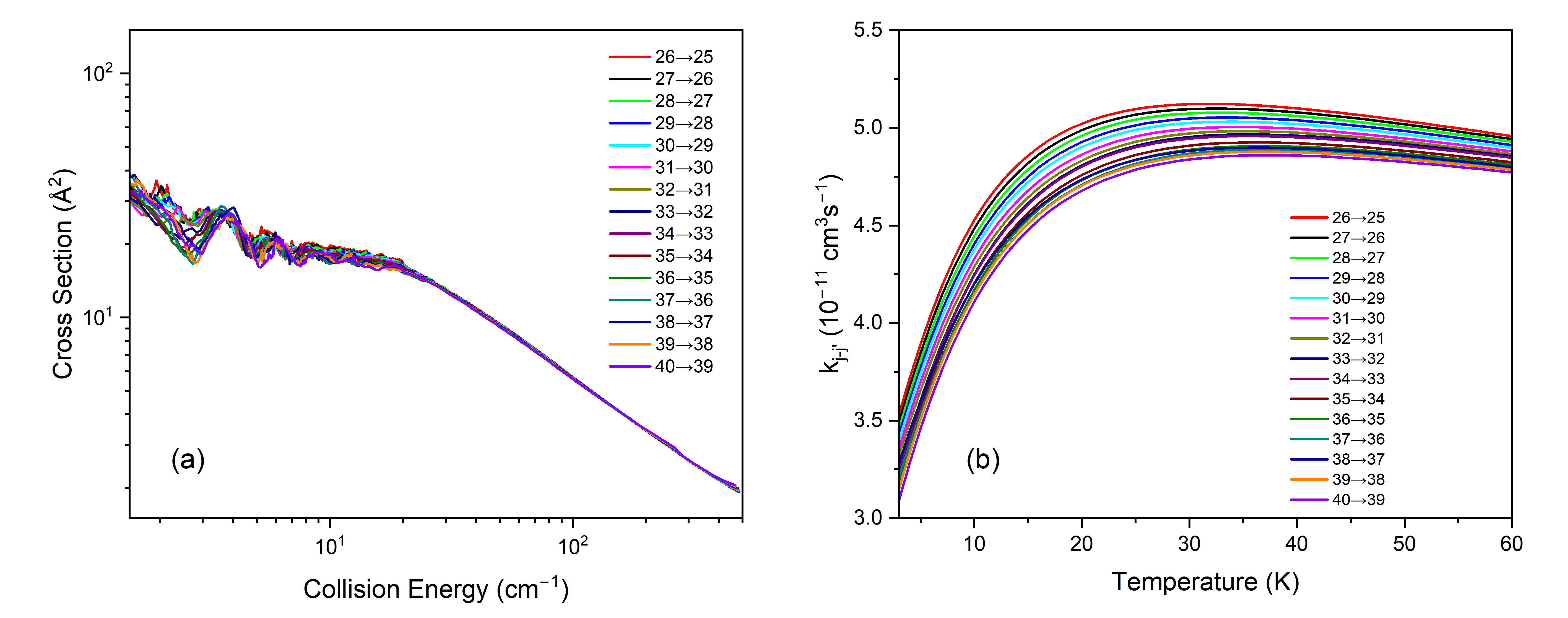}
	\caption{CS cross-sections (a) and rate coefficients (b) for high-$j$ transitions from $26\rightarrow25 \text{ to } 40\rightarrow39$ of SiC$_6$--He.}
	\label{fig.4}
\end{figure}

Following the CC/CS comparison, we present the CS cross sections and thermal rate coefficients for the high-$j$ transitions considered in this study. Fig.~\ref{fig.4}(a) displays the scattering cross-sections for high-lying rotational transitions (from $26\rightarrow25$ up to $40\rightarrow39$) calculated using the CS method. 
In the low-energy region, distinct resonance structures appear in the cross sections for the transitions shown, with slight variations in resonance positions and amplitudes. As collision energy increases, these resonances gradually vanish, and the cross sections generally decrease in the high-energy region. Furthermore, the differences between transitions diminish significantly: the curves nearly overlap for energies above 30\,cm$^{-1}$, and the cross sections decrease to approximately 2\,Å$^2$ at 500\,cm$^{-1}$. This trend indicates that the cross sections of the high-$j$ transitions become increasingly similar at high energies. Fig.~\ref{fig.4}(b) illustrates the temperature dependence of the rate coefficients obtained through thermal averaging of the cross sections. Within the 3--60\,K range, the rate coefficients for all transitions initially rise rapidly with temperature, peak between approximately 25\,K and 30\,K, and subsequently decline slowly. The rate coefficient values for adjacent transitions involving high-$j$ levels are very close; at any given temperature, the rate coefficient is slightly lower for higher initial $j$ values, resulting in only minor offsets between the curves. The CS rate coefficients obtained in this study cover rotational transitions up to $j=40$ over a temperature range of 3--60\,K. Their agreement with the CC results over the tested range and their smooth temperature dependence support the use of CS rate coefficients in non-LTE excitation modelling of SiC$_6$ emission in IRC+10216.

\section{Astrophysical applications}
\label{sec:astro}
\begin{table*}[t]
	\caption{Critical densities $n_{\mathrm{c}}$ ($\mathrm{cm}^{-3}$) of
		$\mathrm{SiC_6}$ at 10, 15, 20, and 50\,K for the observed transitions.}
	\label{tab:1}
	\centering
	\small
	\setlength{\tabcolsep}{6pt}
	\begin{tabular}{lcccc}
		\hline\hline
		$J\rightarrow J'$ & 10\,K & 15\,K & 20\,K & 50\,K \\
		\hline
		$26\rightarrow25$ & $3.7166\times10^{4}$ & $3.4463\times10^{4}$ & $3.2639\times10^{4}$ & $2.7606\times10^{4}$ \\
		$27\rightarrow26$ & $4.1173\times10^{4}$ & $3.8163\times10^{4}$ & $3.6145\times10^{4}$ & $3.0589\times10^{4}$ \\
		$28\rightarrow27$ & $4.5390\times10^{4}$ & $4.2056\times10^{4}$ & $3.9841\times10^{4}$ & $3.3762\times10^{4}$ \\
		$29\rightarrow28$ & $4.9924\times10^{4}$ & $4.6231\times10^{4}$ & $4.3797\times10^{4}$ & $3.7146\times10^{4}$ \\
		$30\rightarrow29$ & $5.4777\times10^{4}$ & $5.0696\times10^{4}$ & $4.8027\times10^{4}$ & $4.0764\times10^{4}$ \\
		$31\rightarrow30$ & $6.0007\times10^{4}$ & $5.5475\times10^{4}$ & $5.2538\times10^{4}$ & $4.4599\times10^{4}$ \\
		$32\rightarrow31$ & $6.5396\times10^{4}$ & $6.0459\times10^{4}$ & $5.7275\times10^{4}$ & $4.8673\times10^{4}$ \\
		$33\rightarrow32$ & $7.1271\times10^{4}$ & $6.5832\times10^{4}$ & $6.2346\times10^{4}$ & $5.2985\times10^{4}$ \\
		$34\rightarrow33$ & $7.7282\times10^{4}$ & $7.1488\times10^{4}$ & $6.7744\times10^{4}$ & $5.7593\times10^{4}$ \\
		$35\rightarrow34$ & $8.3856\times10^{4}$ & $7.7410\times10^{4}$ & $7.3312\times10^{4}$ & $6.2345\times10^{4}$ \\
		$36\rightarrow35$ & $9.0196\times10^{4}$ & $8.3265\times10^{4}$ & $7.8908\times10^{4}$ & $6.7305\times10^{4}$ \\
		$37\rightarrow36$ & $9.7228\times10^{4}$ & $8.9748\times10^{4}$ & $8.5061\times10^{4}$ & $7.2591\times10^{4}$ \\
		$38\rightarrow37$ & $1.0398\times10^{5}$ & $9.6382\times10^{4}$ & $9.1502\times10^{4}$ & $7.8227\times10^{4}$ \\
		$39\rightarrow38$ & $1.1194\times10^{5}$ & $1.0359\times10^{5}$ & $9.8294\times10^{4}$ & $8.4031\times10^{4}$ \\
		$40\rightarrow39$ & $1.2150\times10^{5}$ & $1.1198\times10^{5}$ & $1.0601\times10^{5}$ & $9.0324\times10^{4}$ \\
		\hline
	\end{tabular}
\end{table*}

Since H$_2$ is the dominant collision partner in interstellar molecular clouds, astrophysical modelling requires SiC$_6$--H$_2$ collisional rate coefficients. In the absence of a SiC$_6$--H$_2$ interaction PES, we estimated these coefficients from the calculated SiC$_6$--He rate coefficients using the reduced-mass scaling approximation:
\begin{equation}
	\label{eq:scaling}
	k_{\mathrm{SiC_6-H_2}}(T) =
	\sqrt{\frac{\mu_{\mathrm{SiC_6-He}}}{\mu_{\mathrm{SiC_6-H_2}}}}\,
	k_{\mathrm{SiC_6-He}}(T) \,.
\end{equation}
The resulting scaling factor is 1.396.

This approximation provides a practical estimate but does not account for differences between the SiC$_6$--He and SiC$_6$--H$_2$ interaction potentials. Depending on the transition and molecular species, H$_2$-based rate coefficients may differ from their He-based counterparts by factors of approximately 1--3 \citep{Lique2006,Otoniel2013}. The scaled rate coefficients were therefore used as approximate H$_2$ collision rates in the subsequent non-LTE radiative-transfer calculations.

We performed non-LTE radiative-transfer calculations using RADEX \citep{van2007}. For a representative high-$j$ transition such as $19\rightarrow18$ (The article has not been provided.), our tests show that the $T_{\mathrm{EX}}$ and $T_{\mathrm{B}}$ curves obtained using the CC and CS rate coefficients nearly overlap, with relative deviations within 5\% at $n(\mathrm{H_2}) \ge 2\times10^{3}\,\mathrm{cm}^{-3}$. In contrast, for lower-$j$ transitions such as $7\rightarrow6$, the CC and CS calculations yield visibly different $T_{\mathrm{EX}}$ curves, whereas the corresponding $T_{\mathrm{B}}$ curves remain essentially indistinguishable. These results confirm that the CS approximation is highly reliable for the high-$j$ transitions, justifying its use for the subsequent modelling.

In astrophysical environments, the excitation of molecular rotational levels is governed by the competition between collisional and radiative processes. A commonly used measure of this competition is the critical density \citep{lique2018}, expressed as

\begin{equation}
	\label{eq:ncrit}
	n_{\mathrm{crit}} = \frac{A_{j \to j'-1}}{\sum_{j' < j} k_{j \to j'}}.
\end{equation}

The critical density is defined as the ratio between the Einstein coefficient and the sum of collisional de-excitation rate coefficients. This quantity has been used to assess the validity of LTE for molecular emission in interstellar environments \citep{lique2014,lique2009}. Motivated by the tentative detection of SiC$_6$ toward IRC+10216 reported by \citet{Pardo2025}, we focus on the $j_{\mathrm{up}}=26$--40 rotational transitions covered by their observations. We calculated the Einstein A coefficients for $\Delta j=-1$ rotational transitions with $j$ values ranging from 26 to 40; the corresponding values are available as online data (see Data Availability). The critical densities of these transitions (up to $j=40$), evaluated at $T_{\mathrm{kin}}=10$, 15, 20 and 50 K using the CS rate coefficients, are listed in Table~\ref{tab:1}.

\begin{figure}
	\centering
	\includegraphics[width=\hsize]{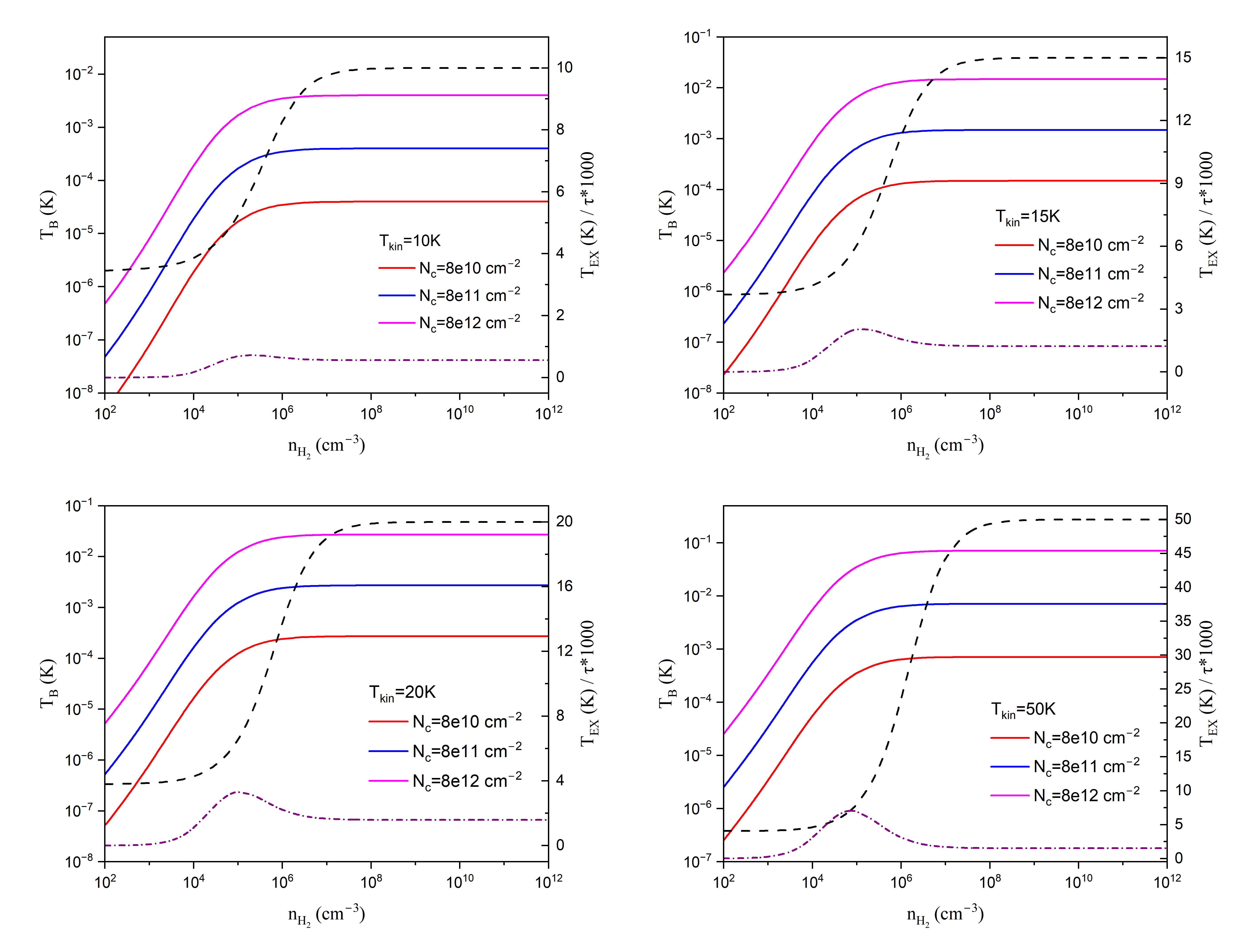}
	\caption{$T_{\mathrm{EX}}$ and $T_{\mathrm{B}}$ for the SiC$_6$ $40\rightarrow39$ transition as functions of the $n(\mathrm{H_2})$ at three SiC$_6$ column densities. Results are based on CS calculations at kinetic temperatures $T_{\mathrm{kin}}=10$, 15, 20 and 50 K. The optical depth is scaled by a factor of 1000.}
	\label{fig.5}
\end{figure}

Fig.~\ref{fig.5} shows the density dependence of $T_{\mathrm{EX}}$ and $T_{\mathrm{B}}$ for the $40\rightarrow39$ transition at $T_{\mathrm{kin}}=10$, 15, 20, and 50\,K. Different colors represent SiC$_6$ column densities of $8.0\times10^{10}$, $8.0\times10^{11}$, and $8.0\times10^{12}$ cm$^{-2}$, respectively. The black dashed curves represent $T_{\mathrm{EX}}$, while the purple dash-dotted curves show the optical depth scaled by a factor of 1000. Consistent with the behaviour of the $19\rightarrow18$ transition, $T_{\mathrm{EX}}$ for the $40\rightarrow39$ transition approaches the cosmic microwave background temperature of 2.73\,K at low densities and begins to increase at $n(\mathrm{H_2})\approx10^{4}$--$10^{5}$\,cm$^{-3}$. This density range is broadly consistent with the critical density of approximately $10^{4}$\,cm$^{-3}$. As the $n(\mathrm{H_2})$ increases further, $T_{\mathrm{EX}}$ approaches $T_{\mathrm{kin}}$. Knowing that when $T_{\mathrm{EX}}$ asymptotically approaches the gas kinetic temperature, which is 10, 15, 20 and 50\,K in our case, LTE is reached. The thermalization of the emission lines indicates the onset of LTE at volume densities $n(\mathrm{H_2}) \gtrsim 10^{6}\,\mathrm{cm}^{-3}$. Meanwhile, $T_{\mathrm{B}}$ rises systematically with increasing column density. The opacities of the detected lines are also displayed in Fig.~\ref{fig.5}. These plots show that the optical depth is less than 0.01 for all emission lines, confirming the optically thin regime.

\begin{figure}
	\centering
	\includegraphics[width=\hsize]{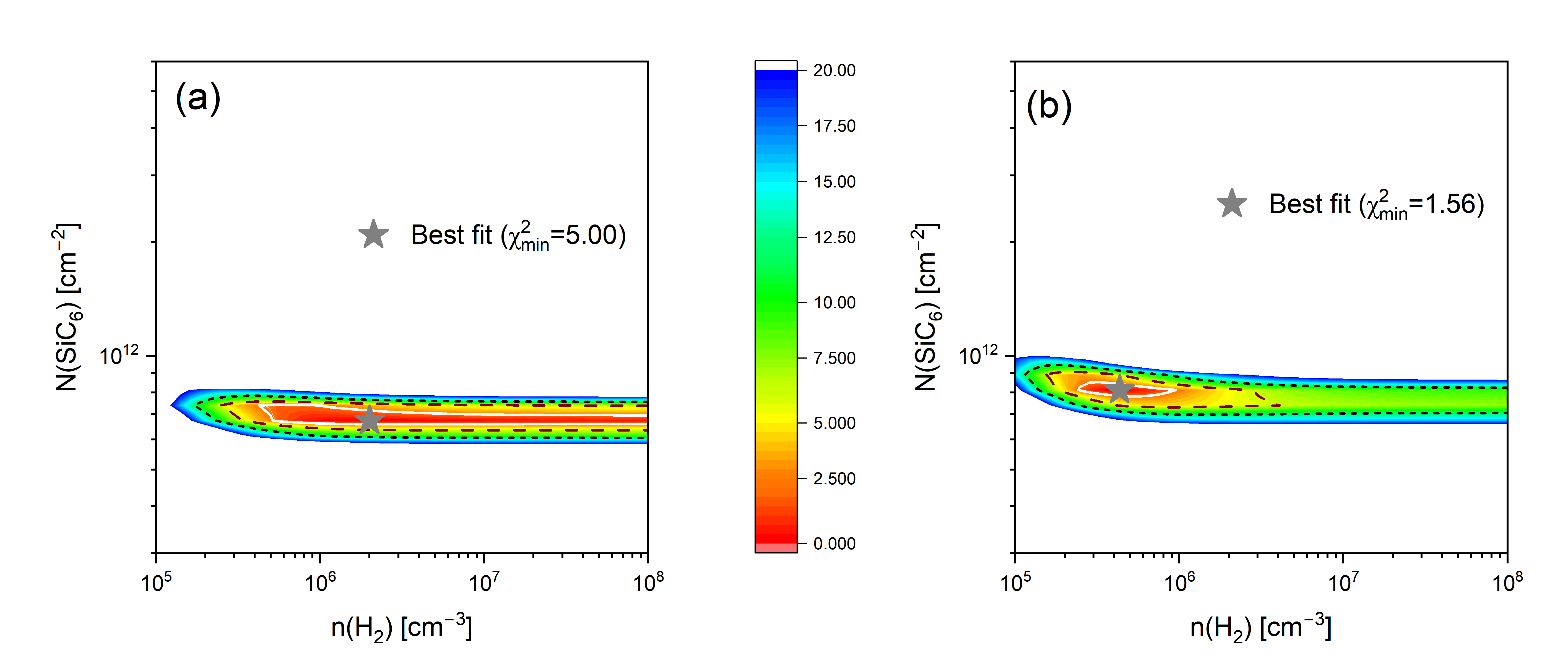}
	\caption{2D $\chi^2$ minimization grids as a function of the  $n(\mathrm{H_2})$ and  $N(\mathrm{SiC_6})$ at $T_{\mathrm{kin}} = 19$~K. Panel (a) includes 9 transitions (including slightly blended features) yielding a best fit of $\chi^2_{\mathrm{min}} = 5.00$. Panel (b) excludes the heavily blended features, using only 7 transitions, yielding $\chi^2_{\mathrm{min}} = 1.56$. The stars mark the best-fit models. Solid white, dashed wine, and short dashed black curves represent represent $1\sigma$, $2\sigma$, and $3\sigma$ levels ($\Delta\chi^2 = 2.30, 6.18, 11.83$), respectively.}
	\label{fig.6}
\end{figure}

\begin{figure}
	\centering
	\includegraphics[width=\hsize]{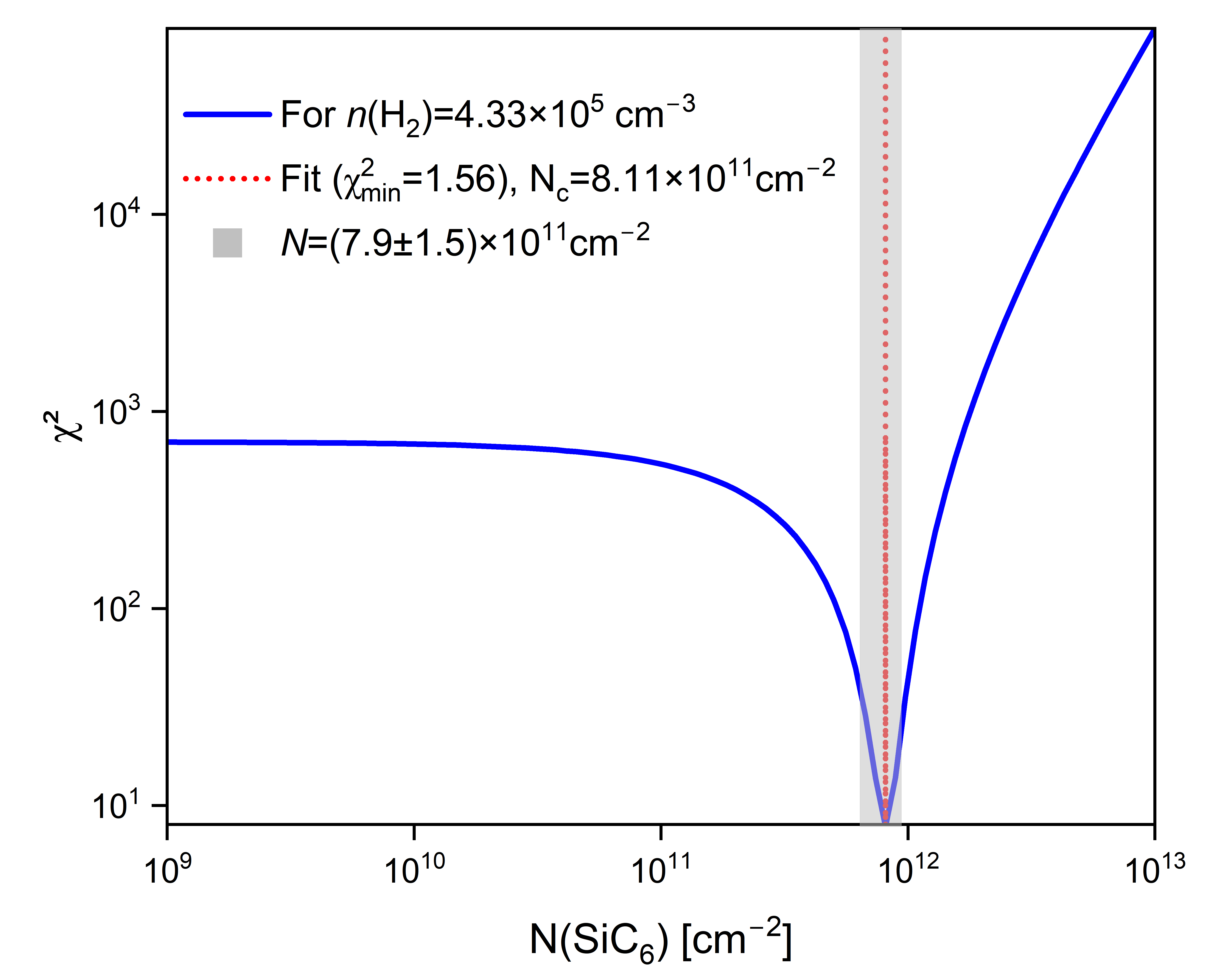}
	\caption{1D $\chi^2$ profile evaluated at the optimal density $n(\mathrm{H_2}) = 4.33 \times 10^5\,\mathrm{cm^{-3}}$ as a function of $N(\mathrm{SiC_6})$. The vertical red short dotted line indicates our optimal non-LTE fit ($N = 8.11 \times 10^{11}\,\mathrm{cm^{-2}}$), while the LT gray shaded area marks the literature value from \citet{Pardo2025}.}
	\label{fig.7}
\end{figure}

Although achieving complete thermal equilibrium requires densities exceeding $10^6 \text{ cm}^{-3}$, the actual physical conditions can be directly constrained by fitting the observed line fluxes of high-$j$ energy levels. To quantitatively assess the impact of non-local thermodynamic equilibrium (non-LTE) effects and to verify the column densities derived from the observed radiation, we performed a $\chi^2$ minimization procedure.
\begin{equation}
	\chi^2 = \sum_{i=1}^{N} \left( \frac{I_{\mathrm{calc},i} - I_{\mathrm{obs},i}}{\sigma_{I,i}} \right)^2
	\label{eq:chi2}
\end{equation}
where $I_{\mathrm{calc},i}$ and $I_{\mathrm{obs},i}$ are the modeled and observed integrated intensities for the $i$-th transition, respectively, and $\sigma_{I,i}$ represents the observational uncertainty.

We compared the observed integrated line intensities reported by \citet{Pardo2025} with those predicted by RADEX, assuming a kinetic temperature of $T_{\mathrm{kin}} = 19\,\mathrm{K}$. Fig.~\ref{fig.6} shows the 2D $\chi^2$ distribution maps to test the robustness of our fitting. Panel (a) includes all 9 transitions in our initial fitting sample, yielding a best fit at $n(\mathrm{H_2}) = 2.01 \times 10^6\,\mathrm{cm^{-3}}$ and $N(\mathrm{SiC_6}) = 6.73 \times 10^{11}\,\mathrm{cm^{-2}}$, with a minimum $\chi^2_{\min} \approx 35.0$ (reduced $\chi^2_{\min} \approx 5$). In this case, the confidence contours are highly elongated, indicating a severe degeneracy between volume and column density. The relatively poor fit suggests that the observed intensities of some transitions are not well reproduced by a single set of model parameters.
In contrast, Fig.~\ref{fig.6}(b) excludes the $J=30\to29$ and $J=36\to35$ transitions, whose observed integrated intensities are 17.271 and 16.118 \,$\mathrm{mK}\cdot\mathrm{km}\cdot\mathrm{s}^{-1}$, respectively, and retains the remaining 7 transitions. Both transitions also deviate from the best-fit linear relation in the rotational-diagram analysis reported by \citet{Pardo2025}. In particular, They noted that the $J=30\to29$ line might be affected by radio-frequency interference (RFI), although to a lesser extent than the excluded $J=32\to31$ line. The consistent deviations of these two transitions in the rotational-diagram analysis and our initial non-LTE fit motivate their exclusion from the subsequent fitting. The origin of the deviation of the $J=36\to35$ transition, however, remains unclear.
The 7-transition dataset yields a significantly improved fit, with $\chi^2_{\min} = 7.81$ (reduced $\chi^2_{\min} \approx 1.56$). The best-fit parameters are $n(\mathrm{H_2}) = 4.33 \times 10^5\,\mathrm{cm^{-3}}$ and $N(\mathrm{SiC_6}) = 8.11 \times 10^{11}\,\mathrm{cm^{-2}}$, and the $1\sigma$ contour is well constrained. Consequently, we adopt the best-fit parameters derived from the 7-transition dataset for the subsequent analysis, while noting that the exclusion of these two transitions is an empirical choice based on the fitting results rather than definitive evidence that the observed intensities are unreliable.

Notably, $n(\mathrm{H_2}) = 4.33 \times 10^5\,\mathrm{cm^{-3}}$, derived from the fit to the 7 selected transitions, is slightly higher than the typical ambient density of the IRC+10216 envelope ($\sim 10^4\,\mathrm{cm^{-3}}$). As shown in Fig.~\ref{fig.5}, at this density, although the $T_{\mathrm{EX}}$ curves for the highest transitions are still in the rising phase, the excitation of the observed transitions appears to approach LTE conditions. The $1\sigma$ and $2\sigma$ confidence contours show a clear degeneracy between volume and column density; however, the optimal column density remains well constrained within the adopted model and fitting procedure.
The column density, $N(\mathrm{SiC_6}) = (7.9 \pm 1.5) \times 10^{11}\,\mathrm{cm^{-2}}$, reported by \citet{Pardo2025} is indicated by the LT gray shaded area in Fig.~\ref{fig.7}. Our non-LTE best-fit value, $N(\mathrm{SiC_6}) = 8.11 \times 10^{11}\,\mathrm{cm^{-2}}$, is in excellent agreement with this estimate and lies well within its reported uncertainty. This agreement suggests that, under the conditions considered here, the non-LTE model yields a column density consistent with the LTE estimate, although it does not necessarily imply complete thermalization of the rotational populations. Therefore, the LTE model can provide a good approximation for deriving the total column density for these observations.

The agreement between the modeled and observed line intensities also provides an observational consistency check for the CS-based modeling. Our RADEX calculations were performed using the CS rate coefficients computed up to $j=40$. The successful reproduction of the observed intensities of the 7 selected transitions and the consistency of the resulting column density with the LTE estimate support the applicability of the CS rate coefficients to modeling the observed high-$j$ SiC$_6$ transitions under the conditions considered here. In summary, our calculations provide collisional data for SiC$_6$ and yield a non-LTE column density for IRC+10216 that is consistent with the previously reported LTE estimate.

\section{Discussion and conclusions}
\label{sec:conclusions}

In this work, state-to-state rotational de-excitation rate coefficients for SiC$_6$--He collisions were calculated using a high-precision PES. Comparisons between the CC and CS methods show that the rate coefficients generally agree within a factor of $\sim 2$, although larger differences occur for some weak transitions and in resonance-dominated regions. Radiative-transfer calculations yield very similar excitation and brightness temperatures for the high-$j$ transitions examined, supporting the use of the CS approximation when fully converged CC calculations are computationally impractical. The CS rate coefficients are provided for transitions involving rotational levels up to $j=40$ over a temperature range of 3--60\,K.

The observed SiC$_6$ emission from IRC\,+10216 was also modelled using RADEX, with SiC$_6$--H$_2$ rate coefficients estimated by reduced-mass scaling of the SiC$_6$--He rate coefficients. A $\chi^2$ analysis of seven selected transitions yields a best-fit of $n(\mathrm{H_2}) = 4.33 \times 10^5\,\mathrm{cm^{-3}}$ and  $N(\mathrm{SiC_6}) = 8.11 \times 10^{11}\,\mathrm{cm^{-2}}$. The derived column density is in good agreement with the value obtained from the observational LTE analysis. These results support the applicability of the CS-based approach to modelling the high-$j$ transitions considered here.
The complete set of state-to-state rate coefficients and the associated RADEX data provide essential collisional input for interpreting interstellar SiC$_6$ observations. These data will be made available through the LAMDA database.

\begin{acknowledgments}
The authors thank Prof. François Lique at the  Institut de Physique de Rennes, Université de Rennes for his helpful discussions. This study was supported by the Scientific Research Development Fund of Zhejiang A\&F University (Grant No.~2024LFR048). 

\facilities{}

\software{MOLPRO \citep{werner2012,werner2020},  
	MOLSCAT \citep{Hutson2019}, 
	RADEX \citep{van2007}
}
\end{acknowledgments}


\section*{Data availability}
The collisional rate coefficients and the associated molecular data required for the RADEX calculations will be deposited in the Zenodo repository and submitted to the LAMDA database upon acceptance for publication.


\bibliography{sample702}{}

@article{Alexander1987,
	author = {Alexander, Millard H. and Manolopoulos, David E.},
	title = {A stable linear reference potential algorithm for solution of the quantum close-coupled equations in molecular scattering theory},
	journal = {JChPh},
	volume = {86},
	number = {4},
	pages = {2044-2050},
	ISSN = {0021-9606
	1089-7690},
	DOI = {10.1063/1.452154},
	year = {1987},
	type = {Journal Article}
}

@article{Bop2025,
	author = {Bop, Cheikh T. and Lique, Fran\c{c}ois},
	title = {Guidelines for non-LTE modelling of cyanopolyynes},
	journal = {A\&A},
	volume = {695},
	ISSN = {0004-6361
	1432-0746},
	DOI = {10.1051/0004-6361/202553698},
	year = {2025},
	type = {Journal Article}
}

@article{Boys1970,
	author = {Boys, S. F. and Bernardi, F.},
	title = {The calculation of small molecular interactions by the differences of separate total energies. Some procedures with reduced errors},
	journal = {MolPh},
	volume = {19},
	number = {4},
	pages = {553-566},
	ISSN = {0026-8976
	1362-3028},
	DOI = {10.1080/00268977000101561},
	year = {1970},
	type = {Journal Article}
}

@article{Cernicharo1989,
	author = {Cernicharo, J. and Gottlieb, C. A. and Guelin, M. and Thaddeus, P. and Vrtilek, J. M.},
	title = {Astronomical and laboratory detection of the SiC radical.},
	journal = {ApJ},
	volume = {341},
	pages = {L25-L28},
	DOI = {10.1086/185449},
	year = {1989},
	type = {Journal Article}
}

@article{Cernicharo2025,
	author = {Cernicharo, J. and Pardo, J. R. and Ag\'undez, M. and Fonfr\'ia, J. P. and Velilla-Prieto, L. and Cabezas, C. and Tercero, B. and de Vicente, P. and Gu\'elin, M.},
	title = {Detection of the linear SiC3 and SiC5 radicals in IRC+10216},
	journal = {A\&A},
	volume = {700},
	ISSN = {0004-6361
	1432-0746},
	DOI = {10.1051/0004-6361/202556370},
	year = {2025},
	type = {Journal Article}
}

@article{Chefai2024,
	author = {Chefai, Atka and Khadri, Fehmi and Alsubaie, Nahaa Eid and Elabidi, Haykel and Hammami, Kamel},
	title = {Collision induced molecular rotation of SiC4--He for astrophysical implications},
	journal = {MNRAS},
	volume = {529},
	number = {4},
	pages = {4066-4072},
	ISSN = {0035-8711
	1365-2966},
	DOI = {10.1093/mnras/stae792},
	year = {2024},
	type = {Journal Article}
}

@article{MacKay1999,
	author = {D. D. S. MacKay and Charnley, S. B.},
	title = {The silicon chemistry of IRC+10216},
	journal = {MNRAS},
	volume = {302},
	pages = {793-800},
	DOI = {10.1046/j.1365-8711.1999.02175.x},
	year = {1999},
	type = {Journal Article}
}

@article{Otoniel2013,
	author = {Denis-Alpizar, Otoniel and Stoecklin, Thierry and Halvick, Philippe and Dubernet, Marie-Lise},
	title = {Rotational relaxation of CS by collision with ortho-and para-H2 molecules},
	journal = {JChPh},
	volume = {139},
	number = {20},
	ISSN = {0021-9606},
	year = {2013},
	type = {Journal Article}
}

@article{Endres2016,
	author = {Endres, Christian P. and Schlemmer, Stephan and Schilke, Peter and Stutzki, J\"urgen and M\"uller, Holger S. P.},
	title = {The Cologne Database for Molecular Spectroscopy, CDMS, in the Virtual Atomic and Molecular Data Centre, VAMDC},
	journal = {JMoSp},
	volume = {327},
	pages = {95-104},
	ISSN = {00222852},
	DOI = {10.1016/j.jms.2016.03.005},
	year = {2016},
	type = {Journal Article}
}

@article{Goldhaber1984,
	author = {Goldhaber, DM and Betz, AL },
	title = {Silane in IRC+ 10216},
	journal = {ApJ},
	volume = {279},
	pages = {L55-L58},
	DOI = {10.1086/184255},
	ISSN = {0004-637X},
	year = {1984},
	type = {Journal Article}
}

@article{Gordon2000,
	author = {Gordon, V. D. and Nathan, E. S. and Apponi, A. J. and McCarthy, M. C. and Thaddeus, P. and Botschwina, P.},
	title = {Structures of the linear silicon carbides SiC4 and SiC6: Isotopic substitution and Ab Initio theory},
	journal = {JChPh},
	volume = {113},
	number = {13},
	pages = {5311-5320},
	ISSN = {0021-9606
	1089-7690},
	DOI = {10.1063/1.1290126},
	year = {2000},
	type = {Journal Article}
}

@article{Green1975,
	author = {Green, Sheldon},
	title = {Rotational excitation in H2--H2 collisions: Close-coupling calculations},
	journal = {JChPh},
	volume = {62},
	number = {6},
	pages = {2271-2277},
	ISSN = {0021-9606
	1089-7690},
	DOI = {10.1063/1.430752},
	year = {1975},
	type = {Journal Article}
}

@article{Hutson2019,
	author = {Hutson, Jeremy M. and Le Sueur, C. Ruth},
	title = {molscat: A program for non-reactive quantum scattering calculations on atomic and molecular collisions},
	journal = {CoPhC},
	volume = {241},
	pages = {9-18},
	ISSN = {00104655},
	DOI = {10.1016/j.cpc.2019.02.014},
	year = {2019},
	type = {Journal Article}
}

@article{Khadri2020,
	author = {Khadri, F. and Chefai, A. and Hammami, K.},
	title = {Low-temperature rate constants and radiative transfer for rotational de-excitation of C5S by collision with He},
	journal = {MNRAS},
	volume = {498},
	number = {4},
	pages = {5159-5165},
	ISSN = {0035-8711
	1365-2966},
	DOI = {10.1093/mnras/staa2746},
	year = {2020},
	type = {Journal Article}
}

@article{Knizia2009,
	author = {Knizia, Gerald and Adler, Thomas B. and Werner, Hans-Joachim},
	title = {Simplified CCSD(T)-F12 methods: Theory and benchmarks},
	journal = {JChPh},
	volume = {130},
	number = {5},
	ISSN = {0021-9606
	1089-7690},
	DOI = {10.1063/1.3054300},
	year = {2009},
	type = {Journal Article}
}

@article{Koelemay2022,
	author = {Koelemay, LA and Burton, MA and Singh, AP and Sheridan, PM and Bernal, JJ and Ziurys, LM},
	title = {Laboratory and astronomical detection of the SiP radical (X2Pi i): more circumstellar phosphorus},
	journal = {Astrophys. J., Lett.},
	volume = {940},
	number = {1},
	pages = {L11},
	DOI = {10.3847/2041-8213/ac9d9b},
	ISSN = {2041-8205},
	year = {2022},
	type = {Journal Article}
}

@article{Gonzalez2024,
	author = {L. Gonz\'alez-S\'anchez and A. Veselinova and A. Mart\'in Santa Dar\'ia and E. Yurtsever and R. Biswas and K. Giri and N. Sathyamurthy and U. Lourderaj and R. Wester and Gianturco, F. A.},
	title = {Collision rate coefficients for C 7 N - and C 10 H - with H 2},
	journal = {ApJ},
	volume = {960},
	pages = {40},
	DOI = {10.1093/mnras/stae2187},
	year = {2024},
	type = {Journal Article}
}

@article{Lique2006,
	author = {Lique, F. and Spielfiedel, A.},
	title = {Ro-vibrational excitation of CS by He},
	journal = {A\&A},
	volume = {462},
	number = {3},
	pages = {1179-1185},
	ISSN = {0004-6361
	1432-0746},
	DOI = {10.1051/0004-6361:20066422},
	year = {2006},
	type = {Journal Article}
}

@article{Manolopoulos1986,
	author = {Manolopoulos, D. E.},
	title = {An improved log derivative method for inelastic scattering},
	journal = {JChPh},
	volume = {85},
	number = {11},
	pages = {6425-6429},
	ISSN = {0021-9606
	1089-7690},
	DOI = {10.1063/1.451472},
	year = {1986},
	type = {Journal Article}
}

@article{McCarthy2000,
	author = {McCarthy, M. C. and Apponi, A. J. and Gottlieb, C. A. and Thaddeus, P.},
	title = {Laboratory Detection of Five New Linear Silicon Carbides: SiC3, SiC5, SiC6, SiC7, and SiC8},
	journal = {ApJ},
	volume = {538},
	pages = {766-772},
	DOI = {10.1086/309177},
	year = {2000},
	type = {Journal Article}
}

@article{McCurdy1977,
	author = {McCurdy, Clyde W. and Miller, William H.},
	title = {Interference effects in rotational state distributions: Propensity and inverse propensity},
	journal = {JChPh},
	volume = {67},
	number = {2},
	pages = {463-468},
	ISSN = {0021-9606
	1089-7690},
	DOI = {10.1063/1.434890},
	year = {1977},
	type = {Journal Article}
}

@article{Mehnen2026,
	author = {Mehnen, Bilel},
	title = {Helium-induced rotational (de-)excitation of interstellar rhomboidal silicon tricarbide (c-SiC3)},
	journal = {A\&A},
	volume = {710},
	ISSN = {0004-6361
	1432-0746},
	DOI = {10.1051/0004-6361/202558085},
	year = {2026},
	type = {Journal Article}
}

@article{Morris1975,
	author = {Morris, M and Gilmore, W and Palmer, Patrick and Turner, BE and Zuckerman, B},
	title = {Detection of interstellar SiS and a study of the IRC+ 10216 molecular envelope},
	journal = {ApJ},
	volume = {199},
	pages = {L47-L51},
	ISSN = {0004-637X},
	DOI = {10.1086/181846},
	year = {1975},
	type = {Journal Article}
}

@article{Ohishi1989,
	author = {Ohishi, Masatoshi and Norio Kaifu and Kentarou Kawaguchi and Akinori Murakami and Shuji Saito and Satoshi Yamamoto and Shin-Ichi Ishikawa and Yoshihisa Fujita and Yutaka Shiratori and Irvine., William M.},
	title = {Detection of a new circumstellar carbon chain molecule, C4Si},
	journal = {ApJ},
	volume = {345},
	pages = {L83-L86},
	DOI = {10.1086/185558},
	year = {1989},
	type = {Journal Article}
}

@article{Pardo2025,
	author = {Pardo, J. R. and Fonfr\'ia, J. P. and Ag\'undez, M. and Tercero, B. and Cabezas, C. and Velilla-Prieto, L. and Gu\'elin, M. and de Vicente, P. and Cernicharo, J.},
	title = {Detection of SiC6 in IRC +10216 through Q-band lines},
	journal = {A\&A},
	volume = {700},
	ISSN = {0004-6361
	1432-0746},
	DOI = {10.1051/0004-6361/202555902},
	year = {2025},
	type = {Journal Article}
}

@article{Sahnoun2018,
	author = {Sahnoun, Emna and Wiesenfeld, Laurent and Hammami, Kamel and Jaidane, Nejmeddine},
	title = {van der Waals interaction of HNCO and H2: Potential Energy Surface and Rotational Energy Transfer},
	journal = {JPCA},
	volume = {122},
	number = {11},
	pages = {3004-3012},
	ISSN = {1089-5639
	1520-5215},
	DOI = {10.1021/acs.jpca.8b00150},
	year = {2018},
	type = {Journal Article}
}

@article{Thaddeus1984,
	author = {Thaddeus, P. and Cummins, S. E. and Linke, R. A.},
	title = {Identification of the SiCC radical toward IC+ 10216-The first molecular ring in an astronomical source. },
	journal = {ApJ},
	volume = {283},
	pages = {L45-L48},
	DOI = {10.1086/184330},
	year = {1984},
	type = {Journal Article}
}

@article{Tian2026,
	author = {Tian, Liman and Ning, Chuangang and Chen, Xiaolin },
	title = {Collisional Excitation of CaO and CaS by Helium: Cross Sections, Rate Coefficients, and Line Ratio Diagnostics},
	journal = {ApJ},
	volume = {1008},
	number = {2},
	ISSN = {0004-637X
	1538-4357},
	DOI = {10.3847/1538-4357/ae9281},
	year = {2026},
	type = {Journal Article}
}

@article{Tian20261,
	author = {Tian, Liman and Sun, Tong and Zou, Wenli and Chen, Xiaolin},
	title = {Rotational (de)-excitation of HNSi and HSiN in collisions with He: ab initio potential energy surfaces, cross-sections, and rate coefficients},
	journal = {MNRAS},
	volume = {545},
	number = {3},
	ISSN = {0035-8711
	1365-2966},
	DOI = {10.1093/mnras/staf2126},
	year = {2026},
	type = {Journal Article}
}

@article{Turner1992,
	author = {Turner, BE.},
	title = {Detection of SiN in IRC+ 10216},
	journal = {ApJ},
	volume = {388},
	pages = {L35-L38},
	ISSN = {0004-637X},
	DOI = {10.1086/186324},
	year = {1992},
	type = {Journal Article}
}

@article{van2007,
	author = {van der Tak, F. F. S. and Black, J. H. and Sch\"oier, F. L. and Jansen, D. J. and van Dishoeck, E. F.},
	title = {A computer program for fast non-LTE analysis of interstellar line spectra},
	journal = {A\&A},
	volume = {468},
	number = {2},
	pages = {627-635},
	ISSN = {0004-6361
	1432-0746},
	DOI = {10.1051/0004-6361:20066820},
	year = {2007},
	type = {Journal Article}
}

@article{Verma1974,
	author = {Verma, R. D. and Nagaraj., S.},
	title = {Absorption Spectrum of the Sic2 Radical in the Gaseous Phase},
	journal = {CaJPh},
	volume = {52},
	pages = {1938-1948},
	DOI = {10.1139/p74-255},
	year = {1974},
	type = {Journal Article}
}

@article{Werner1988,
	author = {Werner, Hans-Joachim and Follmeg, Bernd and Alexander, Millard H.},
	title = {Adiabatic and diabatic potential energy surfaces for collisions of CN(X 2Sigma+, A 2Pi) with He},
	journal = {JChPh},
	volume = {89},
	number = {5},
	pages = {3139-3151},
	ISSN = {0021-9606
	1089-7690},
	DOI = {10.1063/1.454971},
	year = {1988},
	type = {Journal Article}
}

@article{Werner2011,
	author = {Werner, Hans-Joachim and Knowles, Peter J. and Knizia, Gerald and Manby, Frederick R. and Sch\"utz, Martin},
	title = {Molpro: a general-purpose quantum chemistry program package},
	journal = {WIRES. COMPUT. MOL. SCI.},
	volume = {2},
	number = {2},
	pages = {242-253},
	ISSN = {1759-0876
	1759-0884},
	DOI = {10.1002/wcms.82},
	year = {2011},
	type = {Journal Article}
}

@article{Demes2026,
	title={Collisional excitation of cyclopentadiene by helium},
	author={Demes, S{\'a}ndor and Lique, Fran{\c{c}}ois and Ag{\'u}ndez, Marcelino and Cernicharo, Jos{\'e}},
	journal={A \& A},
	DOI = {10.1051/0004-6361/202662015},
	year={2026}
}

@article{lique2018,
	title={Fine-structure transitions of interstellar atomic sulfur and silicon induced by collisions with helium},
	author={Lique, Fran{\c{c}}ois and K{\l}os, J and Le Picard, S{\'e}bastien D},
	journal={PCCP},
	volume={20},
	number={8},
	pages={5427--5434},
	DOI = {10.1039/c7cp05092e},
	year={2018},
	publisher={The Royal Society of Chemistry}
}

@article{lique2014,
	title={Collisional excitation of O 2 by H 2: the validity of LTE models in interpreting O 2 observations},
	author={Lique, F and Kalugina, Y and Chefdeville, S and Van de Meerakker, SYT and Costes, M and Naulin, C},
	journal={A \& A},
	volume={567},
	pages={A22},
	DOI = {10.1051/0004-6361/201423957},
	year={2014},
	publisher={EDP Sciences}
}

@article{lique2009,
	title={The importance of non-LTE models for the interpretation of observations of interstellar NO},
	author={Lique, F and Van Der Tak, FFS and K{\l}os, J and Bulthuis, J and Alexander, MH},
	journal={A \& A},
	volume={493},
	number={2},
	pages={557--563},
	DOI = {10.1051/0004-6361:200810453},
	year={2009},
	publisher={EDP Sciences}
}

@article{werner2012,
	title={Molpro: a general-purpose quantum chemistry program package},
	author={Werner, Hans-Joachim and Knowles, Peter J and Knizia, Gerald and Manby, Frederick R and Sch{\"u}tz, Martin},
	journal={WIRES. Comput. Mol. Sci},
	volume={2},
	number={2},
	pages={242--253},
	DOI = {10.1002/wcms.82},
	year={2012},
	publisher={Wiley Online Library}
}

@article{werner2020,
	title={The Molpro quantum chemistry package},
	author={Werner, Hans-Joachim and Knowles, Peter J and Manby, Frederick R and Black, Joshua A and Doll, Klaus and He{\ss}elmann, Andreas and Kats, Daniel and K{\"o}hn, Andreas and Korona, Tatiana and Kreplin, David A and others},
	journal={JChPh},
	volume={152},
	number={14},
	DOI = {10.1063/5.0005081},
	year={2020},
	publisher={AIP Publishing}
}
\bibliographystyle{aasjournalv7.1}



\end{document}